\documentclass{elsart5p}

\usepackage{graphics}
\usepackage{graphicx}
\usepackage{amssymb}

\newcommand{\bk}{\mathbf k}
\newcommand{\beg}{\begin{equation}}
\newcommand{\en}{\end{equation}}
\newcommand{\dg}{^\dagger}

\begin{document}

\begin{frontmatter}



\title{Superconductivity due to co-operative Kondo effect in Pu 115's}


\author{Maxim Dzero\corauthref{Dzero}},
\ead{dzero@physics.rutgers.edu}
\author{Piers Coleman},

\address{Center for Materials Theory, Serin Physics Laboratory, Rutgers University, Piscataway, NJ 08854}
\corauth[Dzero]{Corresponding author. Tel: (732) 445-4603 fax: (732) 445-4400}

\begin{abstract}
We outline key elements of a theory that accounts for anomalous
properties of the PuCoGa$_5$ and PuRhGa$_5$ compounds as a consequence of a two-body
interference between two Kondo screening channels. 
Virtual valence fluctuations 
of the magnetic Pu configurations create two 
conduction channels of different symmetry.  Using the symplectic-large
N  approach, we are able to demonstrate our pairing mechanism in an
exactly solvable large $N$ limit. 
The critical temperature reaches its maximum 
when the energy levels of excited valence configurations are almost degenerate. 
The symmetry of the order parameter is determined by the product of the 
Wannier form factors in the interfering conduction channels.
\end{abstract}


\end{frontmatter}



Amongst the many heavy fermion superconductors, PuCoGa$_5$ is
extraordinary. 
Besides its unusually high transition
temperature $T_c\simeq 18.5$ K, this material displays local moment
behavior, with a Curie law
susceptibility  right down to the point where superconductivity sets in \cite{Sarrao2002}. Recent NMR relaxation and Knight shift data \cite{Curro2005} 
indicate that superconductivity in this material involves an anisotropic singlet order parameter with 
the lines of nodes at the Fermi surface. Below the superconducting transition temperature
the Knight shift shows no sign of any residual Curie susceptibility, indicating that the local moments
are quenched at the superconducting transition temperature.
Here we outline 
we outline key aspects of a theory of superconductivity in
PuCoGa$_5$ in which the condensation mechanism is actively associated
with the Kondo effect. 
Our theory also 
applies to PuRhGa$_5$, where superconductivity
emerges out of a heavy Fermi liquid state\cite{Wastin2003}. 


The $5f$ electrons of the actinides are less localized than $4f$
electrons of  rare earth materialss, but
more localized than $d$-electrons in transition metals \cite{Sarrao2002}. The intermediate 
nature of $5f$ electrons is a source of an enhanced virtual 
fluctuations between Pu ground $f^5$ and excited $f^4$ and $f^6$ valence configurations,
corresponding to angular momenta $J=5/2$, $J=4$ and $J=0$ respectively.
In a tetragonal crystalline electric field (CEF) environment the 
degeneracies of Pu $J=4$ and $J=5/2$ multiplets are partially
lifted. The symmetry of these multiplets is determined by decomposing the 
representations of the rotation groups
$D_{4}$ and $D_{5/2}$ into irreducible representations of the tetragonal group. 
We assume that the lowest CEF multiplet for $J=5/2$ $f^5$ configuration is a $\Gamma_7$ 
Kramers doublet and for $J=4$ $f^{4}$ configuration the lowest singlet has $\Gamma_3$ symmetry. 
According to a group theoretic analysis, similar to that of Ref. \cite{Cox123},
the hybridization between $f^6$ and $f^5$ is mediated by conduction electrons described 
in the same irreducible representation $\Gamma_7$, while the hybridization between 
$f^4$ and $f^5$ is mediated by conduction electrons described by the 
irreducible representation $\Gamma_6$:
\[
\begin{array}{ccccc}
      &      ~~~\Gamma_6     &       &        ~~\Gamma_7    &   \vspace{-0.3cm}    \\
f^4 & ~\rightleftharpoons & ~f^5 & ~\rightleftharpoons & ~f^6 
\end{array}
\]

Using the Anderson model and integrating out high energy states by means of the Schrieffer-W\"olff
transformation \cite{Cox123} we obtain an effective Hamiltonian:
\beg
\hat{H} = \sum\limits_{\bk\sigma} \epsilon_{{\bf k}\sigma}
c{^{\dag}}_{{\bf k}\sigma} c_{{\bf k}\sigma}
+ \sum_{j }\sum_{\Gamma=1}^{2}J_{\Gamma}{\vec S}_{\Gamma} (j )\cdot \vec{S}_{f} (j),
\label{H}
\en
where $\vec{S_{\Gamma}} (j )= \psi \dg _{\Gamma
} ({\mathbf x}_{j})\vec{S}\psi_{\Gamma } ({\mathbf x}_{j} )$  ($\Gamma=1,2$) is the component of the 
conduction electron spin density with symmetry $\Gamma$ at site $j$, $\psi_\Gamma({\mathbf x}_{j})=
\sum_{\bk}c_{\bk\sigma}\gamma_{\Gamma\bk}e^{-i{\bk}\cdot{\mathbf x}_j}$, 
$\vec{S}_{f} (j)= f\dg _{j\alpha}\vec{S}_{\alpha \beta }f_{j\beta }$ is the spin 
of the localized f-electrons at site j expressed in terms of Abrikosov
pseudofermions and $\gamma_{\Gamma\bk}$ are the Wannier form
factors. Note that conduction electrons in the two channels share a
single Fermi surface and channel symmetry is not conserved when an
electron propagates from site to site. 
It is the violation of channel symmetry it is these processes
that are responsible for an onset of 
superconductivity \cite{CATK,Flint2007}. In fact, there is 
no superconducting instability for the case of a single conduction channel.


The spin $1/2 $ operator $\vec{S}_{f} (j)$ is invariant under charge
conjugation symmetry $f_{\sigma }\rightarrow -\tilde{\sigma}f\dg_{-\sigma }$. 
This gives rise to a continuous $SU (2)$ gauge symmetry 
\[
f_{\sigma}\rightarrow
\cos \theta e^{i\phi } f_{\sigma} + \tilde{ \sigma }\sin \theta 
e^{-i\phi }f\dg _{-\sigma },
\]
which is intimately connected with the Gutzwiller projection
\cite{affleck}. Coleman et al. \cite{CATK} employed this gauge symmetry in the 
context of the two-channel Kondo lattice model (\ref{H}); by factorizing an exchange
interaction (\ref{H}) in terms of spinor $(V_{\Gamma},
\Delta_{\Gamma})$
describting  Kondo hybridization and  pairing respectively, 
they proposed the development of superconductivity with 
the order parameter 
\[
\Lambda\propto|V_1\Delta_2-V_2\Delta_1|,
\]
involving a co-operative interference of the Kondo effect between the two conduction channels. 
Unfortunately,  but the absence of any controlled treatment of this
novel pairing mechanism 
meant that the its validity and application to real systems has remained in doubt. 

Recently, Flint, Dzero and Coleman (FDC) \cite{Flint2007} have
introduced a symplectic large-$N$ approach that links  
time  reversal of  spins with a ``symplectic symmetry''. 
It is the time-reversal symmetry of spins 
that permits electron pairs to form spin-singlets. Conventional large-$N$ theories based
on $SU(N)$ group involve spins that do not correctly time reverse preventing
singlet pair formation. A subset of the generators of the $SU(N)$ group $\{S\}$ 
satisfies a symplectic time-reversal condition
\[
\sigma_2{\vec S}^T\sigma_2\to -\vec S
\]
and by restricting the Kondo interaction to this subset of spins, one
can treat heavy fermion superconductivity in a large $N$ expansion.
The operator expression for these
symplectic spins is
\[
S_{\alpha \beta }= f\dg_{\alpha }f_{\beta }- {\rm sgn} (\alpha ){\rm
sgn} (\beta ) f\dg _{-\beta }f_{-\beta }
\]
where $\alpha, \ \beta  \in [-N/2,N/2]$ and $n_{f}=N/2$.
By restoring the time-reversal of spins, the continuous 
$SU(2)$ gauge symmetry is restored. 
The existence of a well-defined large $N$ limit with a local $SU (2)$
gauge symmetry, means that we now have a firm semi-classical framework for treating this two-component $SU (2)$ spinor, in which $\frac{1}{N}$ plays the role of Planck's
constant, damping out the effect of quantum fluctuations around the
mean-field theory. 


Let us consider the case of $J_1>J_2$. 
In our  symplectic large-$N$ expansion we find that the heavy Fermi liquid forms in channel one below the characteristic 
Kondo temperature $T_{K1}$ followed by the onset of composite pairing 
superconductivity appears below the temperature given by 
\beg
T_c\simeq T_{K1}\exp({-{1}/{\rho_FJ_2}}),
\en
where $\rho_F$ is the density of states at the Fermi level. We suggest this scenario
is realized in the case of PuRhGa$_5$ \cite{Wastin2003}. However, when
the exchange coupling constants are comparable, $J_1\simeq J_2$, the 
composite pairing develops on par with the quenching of magnetic moments
below $T_c\simeq\sqrt{T_{K1}T_{K2}}$. 
Thereby, we find that the highest value of $T_c$ is reached when the energies of the excited
valence configurations are almost degenerate. We believe
that the high value of $T_c$ in PuCoGa$_5$ is governed precisely by that degeneracy,
In both cases, the symmetry of the superconducting gap is given by a product of the Wannier form factors, 
\[
\Lambda_\bk\propto\gamma_{1\bk}\gamma_{2\bk}.
\]  
Our choice of the ground state multiplets
for the Pu $f^5$ and $f^4$ configurations corresponds to the following set of the 
form factors: $\gamma_{1\bk}\propto k_z(k^2-3k_z^2)$ and 
$\gamma_{2\bk}\propto k_z(k_x^2-k_y^2)$. This yields the $d$-wave like symmetry 
of the order parameter in agreement with the earlier suggestions \cite{Curro2005,Opahle2003,Lixt2005}. 

In the single impurity limit of the two-channel Kondo problem there is a quantum critical point
separating two heavy Fermi liquid regimes. From our results it follows 
that in the lattice, quantum critical regime is avoided by Cooper pair formation
via cooperative Kondo effect. To our knowledge this is the first example when a phenomenon of an avoided quantum criticality gives rise to superconductivity. 
 
To summarize, we have discussed a theory for composite pairing in PuCoGa$_5$ and PuRhGa$_5$ 
in the large symplectic-$N$ limit for the two-channel Kondo lattice model. Two screening channels are 
derived from taking into account virtual valence fluctuations of Pu ion. Constructive interference 
between the screening channels in the Kondo lattice leads to the formation of superconducting 
long range order driven by the fomration of composite pairs. The symmetry of the lowest CEF multiplets of corresponding Pu valence configurations define the symmetry of the order parameter. Our work in progress will indicate whether 
an enhancement of the NMR spin relaxation rate just above $T_c$ is a precursor for the channel
interference effect.

This work was supported in part by grant DOE-FE02-00ER45790.


\begin{thebibliography}{99}

\bibitem{Sarrao2002} J. L. Sarrao \emph{et al.}, Nature (London) {\bf 420}, 297 (2002). 

\bibitem{Curro2005} N. J. Curro \emph{et al.}, Nature (London) {\bf 434}, 622 (2005).

\bibitem{Wastin2003} F. Wastin \emph{et al.}, J. Phys. Condens. Matt. {\bf 15}, S2279 (2003).

\bibitem{Opahle2003} I. Opahle and P. M. Oppeneer, Phys. Rev. Lett. {\bf 90}, 157001 (2003).

\bibitem{Lixt2005} L. V. Pourivskii, M. I. Katsnelson and A. I. Lichtenstein, \emph{preprint} cond-mat/0512156.

\bibitem{Cox123} Tae-Suk Kim and D. L. Cox, Phys. Rev. {\bf B55}, 12594 (1997).

\bibitem{CATK} P. Coleman, A. M. Tsvelik, N. Andrei and  H. Y. Kee, Phys. Rev. {\bf B60}, 3608 (1999).

\bibitem{Flint2007} Rebecca Flint, M. Dzero and P. Coleman, \emph{preprint}, (2007).

\bibitem{affleck}I. Affleck, Z. Zou, T. Hsu and P. W. Anderson,
Phys. Rev. {\bf B38}, 745, (1988).

\end{thebibliography}
\end{document}